\documentclass[reprint, aps, prmaterials, amsmath,amssymb,groupedaddress,superscriptaddress]{revtex4-2}

\usepackage{graphicx}
\usepackage{dcolumn}
\usepackage{bm}
\usepackage{color}
\usepackage{enumitem}
\usepackage[normalem]{ulem}
\usepackage{comment}

\newif\ifHighlitedChanges
\def\ifHighlitedChanges{\iftrue}
\ifHighlitedChanges
  
  \def\STRIKE#1{{\color{red}\sout{#1}}}
\else
  
  \def\STRIKE#1{\relax}
\fi

\begin{document}

\title{Data-driven methods for diffusivity prediction in nuclear fuels}

\author{Galen T. Craven}
\email[]{galen.craven@gmail.com}
\affiliation{Theoretical Division, Los Alamos National Laboratory, Los Alamos, New Mexico, USA}
\author{Renai Chen}
\affiliation{Theoretical Division, Los Alamos National Laboratory, Los Alamos, New Mexico, USA}
\author{Michael W. D. Cooper}
\address{Materials Science and Technology Division, Los Alamos National Laboratory, Los Alamos, New Mexico, USA}
\author{Christopher Matthews}
\address{Materials Science and Technology Division, Los Alamos National Laboratory, Los Alamos, New Mexico, USA}
\author{Jason Rizk}
\address{Materials Science and Technology Division, Los Alamos National Laboratory, Los Alamos, New Mexico, USA}
\author{Walter Malone}
\address{Department of Physics, Tuskegee University, Tuskegee, Alabama, USA}
\author{Landon Johnson}
\affiliation{Theoretical Division, Los Alamos National Laboratory, Los Alamos, New Mexico, USA}
\author{Tammie Gibson}
\affiliation{Theoretical Division, Los Alamos National Laboratory, Los Alamos, New Mexico, USA}
\author{David A. Andersson}
\address{Materials Science and Technology Division, Los Alamos National Laboratory, Los Alamos, New Mexico, USA}

\begin{abstract}
The growth rate of structural defects in nuclear fuels under irradiation is intrinsically related to 
the diffusion rates of the defects in the fuel lattice.
The generation and growth of atomistic structural defects can significantly alter the performance characteristics of the fuel. 
This alteration of functionality 
must be accurately captured to qualify a nuclear fuel for use in reactors. 
Predicting the diffusion coefficients of defects and how they impact macroscale properties such as swelling, gas release, and creep is therefore of significant importance in both the design of new nuclear fuels and the assessment of 
current fuel types. 
In this article, we apply data-driven methods focusing on  machine learning (ML) to determine various diffusion properties of two nuclear fuels---uranium oxide and uranium nitride. 
We show that using ML can increase, often significantly, 
the accuracy of predicting diffusivity in nuclear fuels in comparison to current analytical models.
We also illustrate how ML can be used to quickly develop fuel models with parameter dependencies 
that are more complex and robust than what is currently available in the literature.
These results suggest there is potential for ML to accelerate the design, qualification, and implementation of nuclear fuels.
\end{abstract}

\maketitle

\section{Introduction}

Atomistic structural defects influence and alter the macroscopic properties of nuclear fuels and materials \cite{Was2016}. 
Macroscopic changes, such as volumetric swelling, gas release, and creep, 
can in turn give rise to alterations of the functionality of the fuel in a reactor.
Therefore, developing theoretical methods to predict the growth rates of atomistic point defects and defect clusters 
is of significant importance in the design and qualification of fuels and materials that are used in 
reactors.
The growth rates of defect clusters are  governed by the diffusivity of their constituent point defects.
This is because the rate at which defects move through the fuel lattice 
strongly influences how quickly those defects combine to form larger defect clusters. 
Predicting the diffusion properties of defects in nuclear fuels under reactor conditions is, therefore,
an important research focus in reactor design and safety and surety analysis.

There are two primary theoretical approaches that are applied to determine diffusion properties in nuclear fuels: 
(1) deriving empirically-motivated analytical functional forms and fitting those forms to existing experimental data and 
(2) extracting diffusion coefficients from atomic scale calculations and simulations combined with rate theory approaches. 
One simulation method that is commonly applied to understand and predict defect growth in irradiated materials is cluster dynamics \cite{Wolfer1985,Golubov2001,Ortiz2007,Surh2008,Wirth2015,Stewart2018,Kohnert2018}.
Cluster dynamics is a mean-field method that tracks the time evolution of concentrations of point defects and defect clusters \cite{Matthews2019}. 
Diffusivity predictions are generated from cluster dynamics simulation data by combining the predicted defect concentrations with mobility data. 
Empirical analytical models provide ease of use, transferablilty, interpretability, and are computationally simply to evaluate. 
However, they typically have a limited range of applicability with respect to variation of reactor conditions
and often do not capture the salient features of diffusion processes at a quantitative level. 
Atomistic calculations combined with rate models for defect evolution, often under nonequilibrium conditions \cite{craven15c,craven16c}, can be used to provide high-level predictions of diffusion coefficients \cite{Matthews2019,Matthews2020,Zhou2021} in irradiated materials.
However the process to construct, parameterize, calibrate, and test atomistic and cluster dynamics models can be time-consuming.

In this article, we develop a data-driven workflow to predict diffusion properties of nuclear fuels.
This workflow focuses on the application of machine learning (ML) methods  
to predict diffusion coefficients of various chemical species, defect types, and defect clusters.
Machine learning is a broad term that typically defines a set of numerical methods that are applied to construct unknown model functions using existing data to train the ML process \cite{MLbook2, james2013introduction}.
ML methods have had a history of success in the fields of chemistry, physics, and material science \cite{Kulichenko2021review,butler2018machine,Carrasquilla2017,Carleo2017,Biamonte2017quantum,Deng2017,Liu2019}. 
In these fields, ML is often performed by training numerical models through the application of a specific learning algorithm to data from high-level electronic structure calculations or experimental data.
Because ML uses existing data, it represents a powerful compliment to existing methods and not a replacement. 
ML methods can greatly reduce the amount of experimental data or simulation data needed to address a problem,
allowing the prediction of properties of a material without running a computationally-expensive molecular simulation or performing an experiment.
ML also allows for the development of models that capture more complex parameter dependencies than traditional models.

The nuclear engineering/physics community has started to adopt ML methodologies in the study of nuclear reactors and the nuclear fuel cycle~\cite{Morgan2022review,LWR-SNF,Kautz2019,ML_Nucl_1,ML_Nucl_2,Cai2022MLnuclearfuels}. 
Historically, these are research areas in which both experimental and simulation data is scarce and laborious to generate.
Some noteworthy examples of applications of ML in nuclear engineering include: 
the prediction of properties of light water reactors \cite{LWR-SNF},
nuclear criticality safety analysis \cite{ML_Nucl_1}, 
and thermal conductivity prediction~\cite{ML_Nucl_2}. 
Given the previous success of ML in chemistry, physics, material science---and now nuclear engineering--- 
it is promising to expand the utility of ML techniques within the nuclear fuels discipline.

The two specific nuclear fuels examined in this work are uranium oxide $\text{UO}_2$ and uranium nitride $\text{UN}$---the former is an established nuclear fuel and the latter has been not as widely used, although it is a promising candidate for a variety of advanced reactors due to its high thermal conductivity and U density.  \cite{Matthews1988,Chaudri2013,Watkins2021}. 
The diffusion datasets used to train ML processes are obtained from three sources:
experimental results taken from the literature, 
atomistically-informed cluster dynamics simulations, and data augmentation methods that are applied to expand small experimental datasets.
There are a number of recent studies on the diffusion properties of $\text{UO}_2$ \cite{Matthews2019,Matthews2020,Cooper2021,Rest2019review,Perriot2019}.
The data in these studies, and other experimental work \cite{Turnbull1982,MiekeleyFelix1972,DaviesandLong,Sabioni1998}, 
can be used to validate and calibrate the developed models.
There is less available information and data for $\text{UN}$ in comparison to $\text{UO}_2$.
There are, however, several small datasets available for self-diffusion and fission gas diffusion in $\text{UN}$
\cite{Matthews1988,Chaudri2013,Watkins2021,Matzke1989,Matzke1990,HoltandAlmassy,Decrescente1965,Reimann1971,Melehan1964}.
We examine these fuels under irradiated and non-irritated conditions.

\begin{figure}[t]
\centering
\includegraphics[width = 8.5cm]{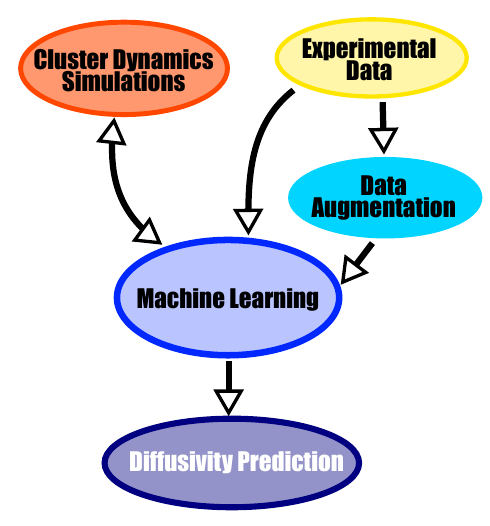}
\caption{\label{fig:TOC}
Schematic representation of the data-driven workflow developed in this work.}
\end{figure}

The primary goal in this paper is to demonstrate the application of ML to predict diffusivity behavior in nuclear fuels. 
The workflow for this project is shown in Fig.~\ref{fig:TOC}.
Experimental diffusion data is used to train ML processes directly and is also fed into data augmentation algorithms to expand small datasets. 
The experimental data and augmented data are coupled with data generated using molecular simulations, specifically cluster dynamics simulations.
The baseline parameters in the cluster dynamics model, informed by atomic scale calculations and simulations, are generated based on DFT and empirical potentials \cite{Cooper2023}.
Those baseline parameters are then calibrated self-consistently by creating a feedback loop between the ML program, the molecular simulations and experimental data used for calibration.
Note the range within which the parameters are allowed to change
is based on the inherent uncertainty of the atomic scale simulations.
The target of this process is to generate diffusivity data from cluster dynamics simulations that agree with experimental results.
The overall output of this workflow is a set of calibrated diffusivity models built by merging data from simulations and experiments.

The remainder of this article is organized as follows:
Section~\ref{sec:methods} contains an introduction to the methods used to examine and predict diffusion coefficients in nuclear fuels.
An overview of the available experimental data is also given.
In Section~\ref{sec:results}, the results of the ML diffusion models are presented.
Section~\ref{sec:uoxide} focuses on fission gas diffusion and self-diffusion in uranium oxide and
Section~\ref{sec:unitride} focuses on various diffusion properties of uranium nitride.
Conclusions and future directions are discussed in Section~\ref{sec:conc}.

\section{Methodologies and Data\label{sec:methods}}

\subsection{Machine Learning \label{sec:ml}}

ML methods are typically applied to construct unknown model functions using existing data to train the ML process \cite{james2013introduction}.
The general objective in {\it supervised} ML, which is the method used here, is to take a collection of input data (sometimes called features) and corresponding output data (sometimes called labels)
and develop a function $f$ that accurately maps the input data to the output data.
The term {\it supervised} ML means that for every set of features $\vec{x}$ there is a corresponding label $y$.
Described in mathematical terms, given a set of $k$ features and a labelled dataset of size $n$, 
the general goal of ML is to take the input data $\bm{X}=\{\vec{x}_1,\vec{x}_2, \ldots, \vec{x}_n\}$ where $\vec{x}_1 = \{x^{(1)}_1,x^{(2)}_1,\ldots, x^{(k)}_1\}$ and corresponding output data $\bm{y}=\{y_1, y_2,\ldots,y_n\}$ and develop a function $f$ 
that accurately maps the input data to the output data:
\begin{equation}
    f: \mathbb{R}^k \to \mathbb{R}.
\end{equation}
In the context of diffusion models developed here, the feature vector $\vec{x}$ will consist of properties such as temperature, partial pressure, and fission rate and the output will be a diffusion coefficient for a specific defect or chemical species.
The goal of training a ML process is to develop a mathematical model using available data so that when 
new data is used as input the model generates an accurate output.
In most ML applications, the developed function $f$ takes a highly complex functional form that does not correspond to a simple and sparse analytical expression.

ML is commonly performed by separating available data into two subsets---training data and testing data. 
The training data is used to build the model, i.e., to train the ML model, and the testing data is used to quantify the predictive accuracy of the model.
The testing data is not used in the model construction and is only used to test and quantify the model.
A typical split between training and testing data is 80:20 meaning that eighty percent of the data is used for training and twenty percent for testing. An 80:20 training/testing ratio is used in all the ML models presented in this work.

There are multiple ML methods that can be used to construct the function $f$. 
Two well-known approaches are  kernel methods and neural networks \cite{MLbook2}.  
Neural networks are data-hungry methods that perform best when 
the size of the dataset used for training is large \cite{Mohan2018}. 
In the context of diffusion in  nuclear fuels, 
the available datasets are typically small and therefore kernel methods 
are expected to perform better for these cases.
The specific ML package we use to perform the ML procedures is {\tt Scikit-learn}~\cite{Scikit-learn}.
In this work, we illustrate data-driven ML methods to determine diffusion of various defects and species in uranium oxide and uranium nitride. The data used to train the ML processes comes from two primary sources: experimental datasets extracted from the literature and molecular simulations. 

\subsection{Experimental Data \label{sec:data}}

The experimental diffusion datasets used as training data for the ML processes are taken from literature sources. 
Table~\ref{tab:table1} is a list of the datasets.
In most cases, the experimental data was given in tabular form in the original papers.
If the numerical values of the data were not listed in a table in the original papers, and were only shown in graphical figures,
the data was digitally extracted from the figures using the \texttt{WebPlotDigitizer} program.
It is important to note that the diffusion data in Table~\ref{tab:table1} was collected over several decades and that 
the experimental techniques used to collect the data are varied, therefore the quality and accuracy of the datasets also vary.

\begin{table*} 
\begin{center}
\caption{List of experimental diffusivity datasets used in this work}   
\label{tab:table1}
\begin{tabular}{|c|c|c|c|c|c|c|}
\hline
Dataset&Fuel&Species& $T$(K)&$p$(atm)&\# of data points&Ref.\\
\hline
Turnbull \textit{et al.}&$\text{UO}_2$&Xe&$\approx500-1700$&$p_{\text{O}_2}$= NA&34&\citenum{Turnbull1982}\\
\hline
Miekeley and Felix&$\text{UO}_2$&Xe&$\approx1200-2000$&$p_{\text{O}_2}$= NA&32&\citenum{MiekeleyFelix1972}\\
\hline
Davies and Long   &$\text{UO}_2$&Xe&fit only & $p_{\text{O}_2}$= NA&fit only&\citenum{DaviesandLong}\\
\hline
Matzke  &$\text{UN}$& N&$\approx1500-2300$ & $p_{\text{N}_2}$= NA&36&\citenum{Matzke1989,Matzke1990} \\
\hline
Holt and Almassy  &$\text{UN}$& N&$\approx2050-2300$&$p_{\text{N}_2}\approx0.01-0.8$&12&\citenum{HoltandAlmassy} \\
\hline
Sturiale and DeCrescente\cite{note1}  &$\text{UN}$& N&$\approx1800-2400$&$p_{\text{N}_2}\approx 0.13$&27&\citenum{Decrescente1965} \\
\hline
Reimann  \textit{et al.}&$\text{UN}$&U&$\approx1875-2150$&$p_{\text{N}_2}\approx10^{-4}-0.6$&30&\citenum{Reimann1971} \\
\hline
\end{tabular}
\end{center}
\end{table*}

\subsection{Cluster Dynamics Simulations\label{sec:simdata}}

Molecular simulation methods can be applied to generate diffusivity data for nuclear fuels \cite{Matthews2019,Matthews2020,Perriot2019}. 
Here, cluster dynamics simulations are used to generate diffusion coefficients for various defect types in the respective fuel.
Implementing the cluster dynamics method consists of parameterizing and then solving a typically large system of 
nonlinear coupled ordinary differential equations, where
each equation in the system describes the time evolution of a 
specific defect type.

The cluster dynamics code {\tt CENTIPEDE} \cite{Matthews2019, Matthews2020} was applied to incorporate physical parameters for the nuclear fuels and to solve the defect evolution equations. 
Details about the {\tt CENTIPEDE} code and the physics behind it can be found in Ref.~\citenum{Matthews2019}.
In a {\tt CENTIPEDE} simulation, the concentration $c_d$ of every defect type $d$ is tracked in time through a differential equation of the form
\begin{equation}
\begin{aligned}
\label{eq:CD}
 \frac{dc_d}{dt}  & = \dot{\beta}_d +\sum_{d'}\dot{R}_{d,d'} (c_d,c_{d'},D_d, D_{d'},T,G) \\
 &\quad-  \sum_{s}\dot{S}_{d,s} (c_d,c_{s},D_d, T,G),  
\end{aligned} 
\end{equation}
where $\dot{\beta}_d$ is the generation rate of defect $d$ due to irradiation,
$\dot{R}_{d,d'}$ is the reaction rate between defect types $d$ and $d'$, 
and $\dot{S}_{d,s}$ is the sink rate between defect type $d$ and sink type $s$.
The sums in Eq.~(\ref{eq:CD}) are taken over all defect types (self-inclusive) and all sink types.
The reaction and sink rates depend on the free energy of the system $G$ and temperature of the system $T$.
The reaction rate between defect types $d$ and $d'$ also depends on the concentrations of each defect and the diffusion coefficients $D_d$ and $D_{d'}$ of those defects.
We are primarily interested in solving for the steady-state concentrations with constant source and sink strengths, which are found
when the rate of change of the concentration vanishes ($\frac{dc_d}{dt} = 0$) up to some numerical precision for all defect types (See Refs~\cite{Matthews2019,Matthews2020,Cooper2023} for more details on {\tt CENTIPEDE} implementations of $\text{UO}_2$). 
Once the concentration of point defects 
and clusters have been determined, 
self-diffusion and Xe diffusion can be obtained 
as the sum over the product of the relative concentration of each defect contributing to diffusion of a species and its mobility.

\section{Results \label{sec:results}}

\subsection{Uranium Dioxide \label{sec:uoxide}}

\begin{figure}[t]
\centering
\includegraphics[width = 8.5cm]{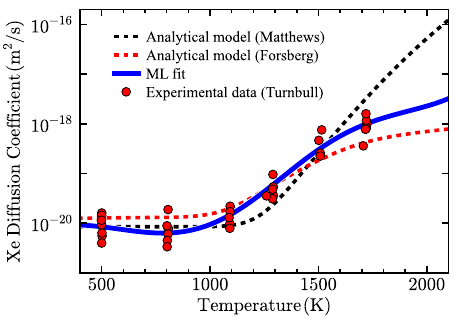}
\caption{\label{fig:Diff_Xe_1}
Xe diffusion coefficient in $\text{UO}_2$ as a function of temperature. The red points are experimental results from Turnbull {\it et al.} \cite{Turnbull1982}. The
solid blue curve is the ML result. The dashed red and dashed black curves are, respectively, 
results of the analytical models of Matthews {\it et al.} \cite{Matthews2020} and Forsberg and Massih \cite{Forsberg2007}.
}
\end{figure}

The generation and subsequent diffusion of fission gas in $\text{UO}_2$ impacts fission gas release and swelling, which, in turn, impact fuel performance   \cite{Matthews2020}.
Turnbull \textit{et al.} have reported measurements for Xe diffusion coefficients in $\text{UO}_2$ under irradiation over the approximate temperature range $500\text{K}-1700\text{K}$ \cite{Turnbull1982}.
Fig.~\ref{fig:Diff_Xe_1} shows the Xe diffusion coefficient in $\text{UO}_2$ 
as a function of temperature---the red circular markers represent the Turnbull data.
Analytical models have been previously developed to predict the Xe diffusion coefficient at various temperatures and under various irradiation conditions. 
Two analytical models are the Matthews model \cite{Matthews2020} and the Forsberg model \cite{Forsberg2007},
which is based on the original development by Turnbull \textit{et al.}~\cite{Turnbull1982}.
The details of these models are found in the Appendix.
In Fig.~\ref{fig:Diff_Xe_1}, the Matthews model is shown as a dashed black curve and the Forsberg model is shown as a dashed red curve.
Both analytical models are in agreement with the Turnbull data over most temperatures, 
but the Matthews model overestimates and the Forsberg model underestimates Xe diffusion at high temperatures.
The parameters in the Matthews model are obtained by fitting to the results of cluster dynamics simulations performed using the CENTIPEDE code, not by directly fitting to the Turnbull data. Therefore, some discrepancy between the data and the model is expected.

As proof of concept that ML can be used to improve the predictive accuracy 
of existing analytical models, 
we used the Turnbull data to train a ML process to predict the Xe diffusion coefficient.
The specific ML method used was kernel ridge regression (KRR), 
a method which
combines ridge regression with the so-called kernel trick\cite{Mohri2018book}. 
We implemented KRR using the {\tt Scikit-learn}~\cite{Scikit-learn} software package. 
Ridge regression is a method for approximating the coefficients of several multiple-regression models which works well if the independent variables are highly correlated~\cite{RidgeRegression1, RidgeRegression2}.
{\tt Scikit-learn} specifically utilizes ridge regression with linear least squares with $L_2$-norm regularization. 
In some ML methods and applications, 
raw data must be transformed via a feature map into a feature vector representation. 
Kernel methods, through the use of kernel functions, can circumvent the need to directly calculate feature vectors, which can be computationally expensive. 
These methods achieve this by calculating the inner products between each image pair within the feature space~\cite{Theodoridis2018book}.

The predictor function (the unknown function to be approximated) in KRR
can be expressed as:
\begin{equation}
    \label{eq:KRR}
    f(\vec{x})=\sum_i\alpha_i\mathcal{K}(\vec{x}, \vec{x}_i),
\end{equation}
where $\mathcal{K}(\vec{x}, \vec{x}_i)$ is the so-called kernel which can be chosen to take various functional forms
depending on the properties of the data being analyzed;
a polynomial kernel and radial basis function kernel~\cite{murphy2012machine}
are used in this work.
The elements $\alpha_i$ in Eq.~\ref{eq:KRR} are taken from the matrix,
\begin{equation}
    \boldsymbol{\alpha}=(\boldsymbol{K} + \lambda\boldsymbol{I})^{-1}\boldsymbol{y},
\end{equation}
where $\boldsymbol{K}$ is the kernel matrix with elements given by $K_{i,j} = \mathcal{K}(\vec{x}_i, \vec{x}_j)$,
$\boldsymbol{I}$ is the identity matrix,
$\lambda$ denotes a regularization parameter
in the ridge regression method that puts a penalty on the weights in order to reduce the variance of predictions,
and $\boldsymbol{y}$ represents the target matrix.

The ML result for Xe diffusivity generated by training a KRR process is shown as a blue curve in Fig.~\ref{fig:Diff_Xe_1}.
The ML result is in strong agreement with the experimental results over the entire temperature range of the Turnbull data.
The dip in the ML model at $\approx 800\text{K}$ is a result of variance in the data used to train the ML model, not a change in the diffusion mechanism.
This proof of concept example illustrates the power of ML to quickly generate models that accurately capture important trends in diffusion data. 
When a ML model is trained using the Turnbull data, there is large variance in the model depending on which points are randomly assigned as testing data and training data. To mitigate this variance, we developed a data augmentation approach to artificially expand the Turnbull dataset, and, therefore, reduce variance in the ML model. 
The results presented in Fig.~\ref{fig:Diff_Xe_1} use this data augmentation method, which is described in detail below.

\begin{figure}[t]
\centering
\includegraphics[width = 8.5cm]{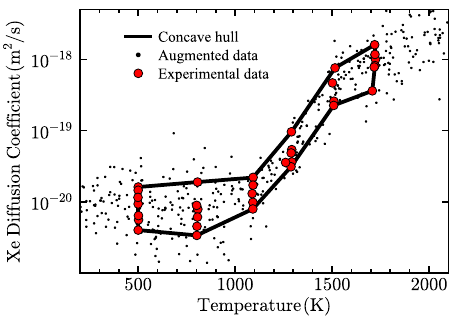}
\caption{\label{fig:Diff_Hull}
The concave hull of the Turnbull {\it et al.} data is shown in solid black. 
The black dots are augmented data and the  
red points are the Turnbull experimental data.
}
\end{figure}

Large datasets are typically used to train ML models (e.g., neural network models).  
This highlights the traditional connection between big data and machine learning.
However, in context of diffusion properties of nuclear fuels, 
there is often limited experimental data available. 
In the situations in which the available data is limited, 
performing ML can result in models with large variances and uncertainties.
To mitigate these problems, 
a data augmentation technique has been developed to artificially expand the small available datasets.
This approach shares similarities with density estimation methods~\cite{tanner2012tools} 
but is more specifically tailored to solve the problem
of fission gas diffusivity in nuclear fuels.
The core of this approach is to use the geometric size of the experimental dataset used as training data to estimate the variance of the data and the density of the experimental data to estimate the mean of the data, and then to generate augmented data by
randomly sampling new datapoints from a distribution that uses the estimated variance and mean.

The first step in the augmentation process is to construct the concave hull (a boundary over the dataset) of the data. 
The concave hull of the Turnbull data is shown in Fig.~\ref{fig:Diff_Hull}.
The Turnbull data consists of diffusivity data  as a function of temperature---the concave hull of the data $\partial\mathcal{H}$ bounds the data.
At a specific temperature $T$, augmented datapoints are generated
by sampling from a normal distribution $\mathcal{N}(\mu(T),\mathcal{W}(T))$ where 
$\mu(T)$ is the distribution mean which is extracted from the hull density and $\mathcal{W}(T)$ is the width of the hull.
The hull width is defined by drawing a vertical line at temperature $T$,
and noting that 
the line intersects the hull boundary twice, once at a higher diffusion value $\partial \mathcal{H}_\text{H}(T)$ and 
once at a lower diffusion value $\partial \mathcal{H}_\text{L}(T)$. The width is $\mathcal{W}(T) = \partial \mathcal{H}_\text{H}(T)- \partial \mathcal{H}_\text{L}(T)$.
The mean is constructed $\mu(T)$ from a spline interpolation across the data.

To generate an augmented datapoint, a random temperature is sampled from a uniform distribution over the
desired temperature range, here $300\text{K}-2100\text{K}$.
At the sampled temperature, a sample is drawn from the normal distribution $\mathcal{N}(\mu(T),\mathcal{W}(T))$.
That sample is an augmented diffusion value at the specified temperature.
We then iterate over this process until the desired number of augmented datapoints are generated.
After performing data augmentation and training the ML model on the augmented data,
the variance in the model development is reduced.
The black dots in Fig.~\ref{fig:Diff_Hull} are augmented data. 
When the variance in the Turnbull data is small, the variance in the augmented data is also small. 
Similarly, when the variance in the data is large, the variance in the augmented data is also large. 
This illustrates how developed data augmentation method captures trends in the variance of the training data.

\begin{figure}[t]
\centering
\includegraphics[width = 8.5cm]{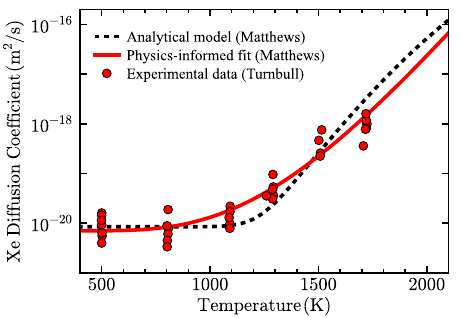}
\caption{\label{fig:Diff_PIML}
Xe diffusion coefficient in $\text{UO}_2$. The red points are experimental results.
The dashed black curve is the analytical model of Matthews {\it et al.} \cite{Matthews2020} and solid red curve is Matthews model parameterized using augmented data. Note that the Matthews model is not fit directly to the Tunbull data, so, here, the primary conclusion is the utility of PIML in the context of fitting to new data sources.
}
\end{figure}

The kernel-based ML approach can be applied to generate accurate models for diffusion data,
but it does not explicitly encode any fundamental physics in the fitting process.
Explicitly encoding physics into ML models can be advantageous
for situations in which the ML model is used to make predictions outside of the boundaries of the training data, i.e., when the ML is used for extrapolation.
One way that physics principles can be encoded into an ML model is to use Physics-Informed Machine Leaning (PIML) methods \cite{Karniadakis2021physics}.
Fig.~\ref{fig:Diff_PIML} shows the result of a PIML model
that is developed by reparameterizing the Matthews analytical model
using augmented data generated from the Turnbull dataset.
The PIML agrees well with the experimental data across all temperatures.
One advantage of the PIML method over the kernel-based ML model is that it captures both the 
high and low temperature trends outside of the boundaries of the experimental data, similar to the Matthews analytical model \cite{Matthews2020}.

\begin{figure}[t]
\centering
\includegraphics[width = 8.5cm]{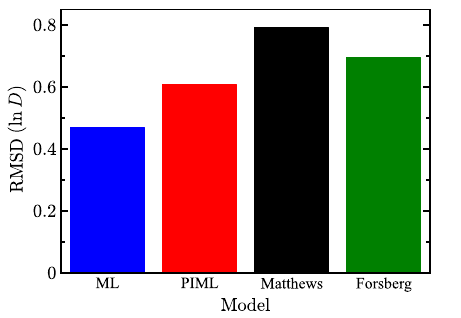}
\caption{\label{fig:ML_err_UO2}
Root mean square deviation of various models for the Xe diffusion coefficient in $\text{UO}_2$. 
The experimental data used to calculate the RMSD is from Turnbull {\it et al.} \cite{Turnbull1982}.
}
\end{figure}

The accuracy of ML models can be compared with the analytical models to quantify any 
improvement in predictive accuracy that can be gained using data-driven methods.
The root mean square deviation (RMSD) for the four methods/models described previously (taken with respect to the Turnbull data) is shown in Fig.~\ref{fig:ML_err_UO2}. 
The ML result gives the lowest error and the PIML gives the second lowest error. 
Both data-driven approaches increase the predictive accuracy for Xe diffusion in comparison to analytical models.
Compared to the Matthews model and the Forsberg model, 
the kernel-based ML approach reduces the error by approximate factors 1.7 and 1.5, respectively.
The PIML method reduces the error by a factor 1.3 in comparison to the Matthews model 
and a factor 1.2 in comparison to the Forsberg model.
Note that all the models including the analytical models perform well for Xe diffusion in $\text{UO}_2$.
This can be observed qualitatively in Figs.~\ref{fig:Diff_Xe_1} and \ref{fig:Diff_PIML}.
Therefore, while the data-driven models do provide improvements in predictive accuracy 
for well-studied fuels such as $\text{UO}_2$, 
we anticipate the primary use of these methods will be when modeling lesser-studied fuels.

\subsubsection{Multi-Dimensional Diffusion Models}

\begin{figure}[t]
\centering
\includegraphics[width = 8.5cm]{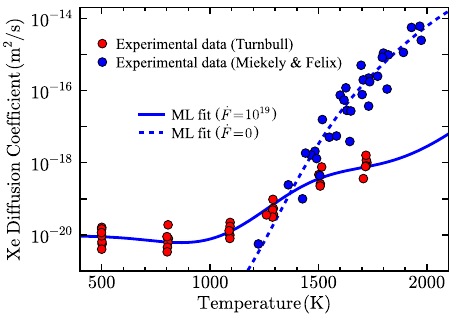}
\caption{\label{fig:Diff_DFT}
Xe diffusion coefficient in $\text{UO}_2$ as a function of temperature $T$ and fission rate $\dot{F}$ for the ML model. The solid curve is the result of the ML model for the irradiated case ($\dot{F}= 10^{19} \, \text{fissions} / \text{m}^3 \, \text{s}$) and the dashed curve is the result for the nonirradiated case ($\dot{F}= 0$). The red points are experimental results from Turnbull {\it et al.} \cite{Turnbull1982} and the blue points are experimental results from Miekeley and Felix \cite{MiekeleyFelix1972}. 
}
\end{figure}

\begin{figure}[t]
\centering
\includegraphics[width = 8.5cm]{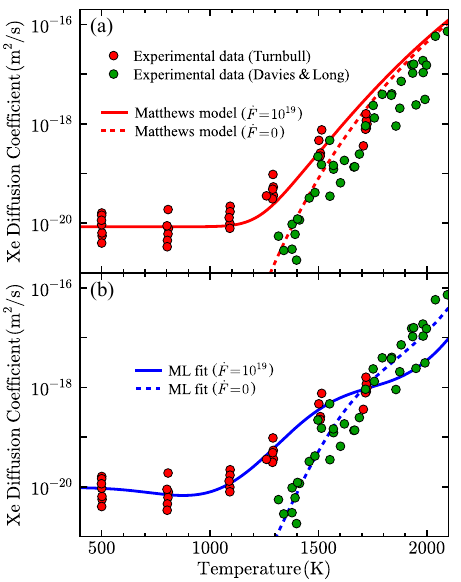}
\caption{\label{fig:Diff_DFT_DL}
Xe diffusion coefficient in $\text{UO}_2$ as a function of temperature $T$ and fission rate $\dot{F}$ for (a) the Matthews model (b) a ML model. The solid curve in each panel is result of the respective model for the irradiated case ($\dot{F}= 10^{19} \, \text{fissions} / \text{m}^3 \, \text{s}$) and the dashed curve is the result for the nonirradiated case ($\dot{F}= 0$). The red points are experimental results from Turnbull {\it et al.} \cite{Turnbull1982} and the green points are experimental results from Davies and Long \cite{DaviesandLong} generated through random sampling inside the error bounds of the fit. 
}
\end{figure}

ML can also be applied to construct multidimensional diffusion models of the form $D(T,\dot{F})$ which capture the dependence of Xe diffusion on the fission rate $\dot{F}$ in the material in addition to capturing trends in the temperature dependence.
Fig.~\ref{fig:Diff_DFT} shows the results of a KRR ML process in comparison
to the irradiated Turnbull data and the 
nonirradiated ($\dot{F}=0$) data of Miekeley and Felix \cite{MiekeleyFelix1972}.  
The ML model is trained on augmented data generated from these datasets.
The ML prediction is in strong agreement with the experimental datasets over all examined temperatures 
and accurately captures the irradiation behavior.
This illustrates the ability of ML to accurately capture trends in multidimensional diffusion data.
We have confirmed that the ML model smoothly interpolates between the fission rate of the Turnbull data ($\dot{F}= 10^{19} \, \text{fissions} / \text{m}^3 \, \text{s}$) and the nonirradiated ($\dot{F}=0$) limit.
However, because experimental data has not to our knowledge been generated at intermediate values between these limits, the ML model may not scale accurately in the intermediate regime due to lack of training data. Calibration of the ML model at intermediate fission rates could be accomplished using data generated from cluster dynamics calculations, for example, by using  {\tt CENTIPEDE}.

Other experimental datasets for Xe diffusion in $\text{UO}_2$  besides the Miekeley and Felix data can be found in the literature.
For example, the analytical fit of Davies and Long is generally considered
to more accurately capture Xe diffusivity at thermal equilibrium. 
Note that the primary goal in this work is to illustrate the utility of ML in nuclear fuel model development,
not to assess the accuracy and validity of the datasets that are available in the literature. 
So, in the context of this work, the datasets are primarily tools to benchmark the developed ML methods.
Shown in Fig.~\ref{fig:Diff_DFT_DL}(a) is a comparison between the results of Matthews model (which was developed to agree with {\tt CENTIPEDE} predictions that are close but not identical to the Davies and Long and Turnbull data sets), the Turnbull data, and the data generated using the analytical fit of Davies and Long. 
The Matthews model is in agreement with both datasets, but overestimates the diffusivity of the nonirradiated data.
For comparison, the ML result shown in Fig.~\ref{fig:Diff_DFT_DL} (b) 
is in excellent agreement with both data sets over all irradiation and temperature conditions.  
Note that the different experimental data sets used in this subsection have different partial oxygen pressures as well as different irradiation conditions.

\begin{figure*}[t]
\centering
\includegraphics[width = 17.0cm]{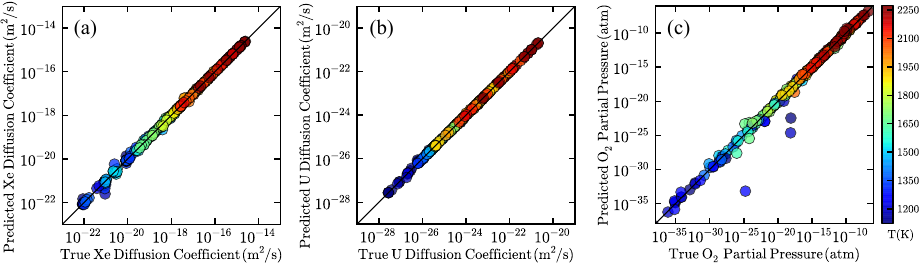}
\caption{\label{fig:Centipede_UO2}
Machine learning results for $\text{UO}_2$ showing predicted vs. true values for (a) Xe diffusion, (b) U diffusion, and 
(c) the partial pressure of $\text{O}_2$. 
The data used to train the models is obtained from \textit{CENTIPEDE} cluster dynamics simulations. 
The diagonal line in each panel illustrates where the predicted value equals the true value.
Different color markers correspond to different temperatures shown in the colorbar to the right.
}
\end{figure*}

Using the results of {\tt CENTIPEDE} cluster dynamics simulations, 
multi-dimensional diffusion models $D(T,\dot{F},p_{\text{O}_2})$ can be developed using ML that include the dependence on the partial pressure of oxygen $p_{\text{O}_2}$ as well as the fission rate and temperature.
The training data for the ML process was generated by 
performing {\tt CENTIPEDE} simulations at 1000 datapoints on a grid in the $\{T,\dot{F}, p_{\text{O}_2}\}$ parameter space.
An 80:20 split between training and testing data was used.
All features and labels were log-scaled before training and testing except the temperature.
To predict the Xe diffusion and U diffusion, we again utilized KRR~\cite{Mohri2018book} 
with the addition of the nearest neighbors (NN) approach \cite{craven20b,craven20c}. 
The procedure is performed by taking an input state point $p = \{T,\dot{F}, p_{\text{O}_2}\}$ (the point where the diffusion coefficient would like to be predicted) and determining the $N_\text{neigh} = 8$ nearest neighbor points in the training data to the input point.
The metric used to determine the nearest neighbors to the input point $p$ was the weighted Euclidean distance
with weights obtained using a grid search hyperparameter optimization.
After the nearest neighbors are determined, the KRR was performed using only the NN points.
The output of the KRR procedure at the target datapoint $p$ in state space is the predicted diffusion value.

Fig.~\ref{fig:Centipede_UO2}(a) and (b) are plots of predicted values vs. true values for
Xe diffusion and U diffusion, respectively.
The average percent error of the testing data was approximately $10\%$ for both U and Xe diffusion, 
illustrating excellent agreement between the ML model and the testing data. 
The different color points in each plot signify different temperatures. 
The data spans the temperature range 1200K to 2250K, and excellent agreement is observed between the ML model and the testing
data over the entire temperature range.
The data spans the fission rate range $10^{17}\, \text{fissions} / \text{m}^3 \, \text{s}$ to $10^{19}\, \text{fissions} / \text{m}^3 \, \text{s}$.

We also used a similar ML procedure to develop a model for the partial pressure of oxygen $p_{\text{O}_2}(T,\dot{F},D)$
that takes diffusivity as an input in addition to fission rate and the temperature.
The result of this procedure is shown in 
Fig.~\ref{fig:Centipede_UO2}(c). 
Excellent agreement is observed between the predicted values and the true values for the partial pressure.
Experimental values for diffusivity are commonly reported at a specific temperature and fission rate.
Because the developed partial pressure model takes typically reported quantities as inputs,
it allows the determination of the thermodynamic state of a fuel or experiment (qualitatively described using the partial pressure, which controls the $x$ in $\text{UO}_{2\pm x}$ linking to the defect concentration).

\subsection{Uranium Nitride \label{sec:unitride}}

The diffusion properties of UN are less studied and less understood than the diffusion properties of $\text{UO}_2$.
This provides an opportunity to illustrate the utility and predictive power of data-driven methods in the context of developing
diffusion models for emerging nuclear fuel candidates.
Training data for diffusivity prediction in UN can be taken from existing experimental datasets, analytical models, new data generated using cluster dynamics simulations, or a combination of these data sources.
Given that the experimental and analytical data are limited and uncertain, the use of cluster dynamics informed by atomic scale simulations is key to providing input for the ML models. 
A list of the UN experimental diffusivity datasets we use is shown in Table~\ref{tab:table1}.
The analytical model we compare to our ML model is taken from a collection of analytical models for diffusivity in UN developed by Hayes~\cite{hayes1990material}.
The details of this model are found in the Appendix.

\begin{figure}[t]
\centering
\includegraphics[width = 8.5cm]{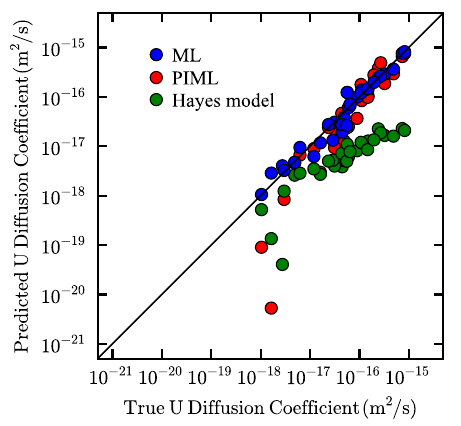}
\caption{\label{fig:true_vs_pred}
Predicted vs. true values for uranium diffusion in $\text{UN}$. Results are shown for a ML model (blue), a PIML model (red), and the Hayes analytical model (green).
The data used as true values are from Reimann {\it et al.} \cite{Reimann1971}.
}
\end{figure}

Fig.~\ref{fig:true_vs_pred} is a plot of true vs. predicted values for U diffusion in UN 
calculated using three different models: the Hayes analytical model, a PIML model, and a kernel-based ML model.
The true values are taken from the experimental measurements of Reimann {\it et al.} \cite{Reimann1971}
which give U diffusion coefficients as a function of temperature $T$ and the partial pressure of nitrogen $p_{\text{N}_2}$.
All of the models we apply are multidimensional diffusion models of the general form  $D(T,p_{\text{N}_2})$ that take
temperature and pressure as inputs and return a predicted U diffusion coefficient as an output.
The Hayes analytical model systematically underestimates the U diffusion values 
and also generates several points with significant error.
A PIML method, developed by reparameterizing the Hayes model, 
improves on the Hayes fit and performs well for most datapoints, 
but there are some points (for lower coefficients end particularly) in which a significant difference exists between 
the true and predicted results
because the reparameterization does not change
the oversimplified construction of the empirical Hayes model.
The ML results give the best results in terms of accuracy and variance.

\begin{figure}[t]
\centering
\includegraphics[width = 8.5cm]{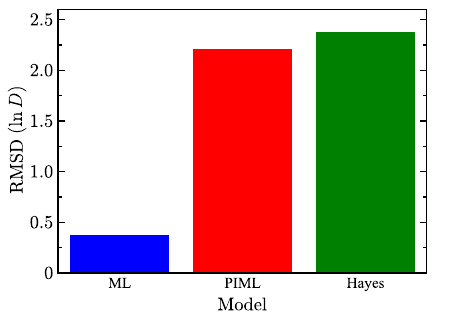}
\caption{\label{fig:ML_err_UN}
Root mean square deviation of various models for the uranium diffusion coefficient in $\text{UN}$. 
The experimental data used to calculate the RMSD is from Reimann {\it et al.} \cite{Reimann1971}.
}
\end{figure}

Shown in Fig.~\ref{fig:ML_err_UN} is a comparison between the error values for each method taken with respect to the experimental data from Reimann \cite{Reimann1971}.
The error is quantified using the RMSD.  
The kernel-based ML method reduces the error by a factor of approximately 6 in comparison to both the PIML method and Hayes analytical model~\cite{hayes1990material}.
This significant error reduction highlights the utility of ML methods for quickly developing accurate diffusion models.

\begin{figure}[t]
\centering
\includegraphics[width = 8.5cm]{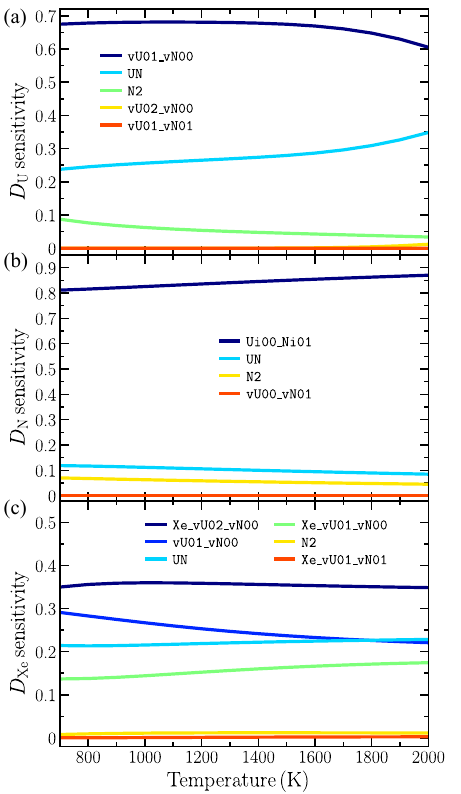}
\caption{\label{fig:middle_SA_stack}
Sensitivity indices for various defect types in UN. Results are shown for (a) U diffusion,  (b) N diffusion, and (c) Xe diffusion.
}
\end{figure}
\subsubsection{Sensitivity Analysis of \text{UN} Diffusion}
There is a limited amount of experimental diffusion data available for UN.
Therefore, mechanistic cluster dynamics models can (and in general must)
be used to augment the experimental data in order to make accurate predictions about diffusivity. 
In cluster dynamics models, understanding which defect types contribute the most to the overall diffusion mechanism of a species
can be quantified and understood using Sensitivity Analysis (SA)---a collection of mathematical and statistical methodologies that are used to understand what inputs contribute the most to a model output.
Sensitivity Analysis is a powerful mathematical tool for building and testing the complex computational 
models that play a significant role in almost all social and physical scientific disciplines.
Some example uses of SA methods in the context of model construction are: (a) identifying the most influential inputs to a model output, (b) improving the understanding of the relations between the inputs and output of a model, (c) calibrating input errors, and (d) assessing the quality and confidence of a model \cite{Iooss2015}.

In this article, we use Global Sensitivity Analysis (GSA) methods 
to investigate the impact of different parameters in the cluster dynamics models for diffusivity. 
Some of the advantages of GSA methods are that
they are capable of handling and analyzing high-dimensional inputs in a computationally efficient manner
and that they can be built to taken into account nonlinear effects in the model \cite{saltelli2008global}.
Specifically, we use the well-known Morris method~\cite{morris1991factorial,campolongo2007effective}
to quantify what defects and parameters are most important for diffusivity in UN. 
The result of applying the Morris method to a model is a collection of sensitivity indices, one index for each input. 
The value of each index quantifies the relative importance of the corresponding input.
A larger value for a sensitivity index implies a higher importance in the model output. 
The Morris method is computationally more efficient than other 
SA methods because it scales linearly with the number input parameters.
Therefore, it is well-suited for analyzing cluster dynamics simulations of UN which involve a high ($>50$) number of input variables.
All the numerical algorithms used to perform SA are taken from the \texttt{SALib} package \cite{Herman2017} 
and implemented using PYTHON scripts.

In our SA results, a sensitivity index is assigned to every defect type that is tracked in the cluster dynamics simulation.
The value of each index quantifies the importance of the corresponding defect to the examined diffusion process. 
SA was performed for N, U, and Xe diffusivity in UN.
The notation \texttt{vU}$x$\texttt{\_}\texttt{vN}$y$ denotes a vacancy cluster of $x$ uranium vacancies and $y$ nitrogen vacancies. A similar notation is used for interstitials. UN represents perfect UN without defects and $\text{N}_2$ nitrogen gas. Crystal \texttt{Xe}\texttt{\_}\texttt{vU}$x$\texttt{\_}\texttt{vN}$y$ denotes that a Xe atom resides in the \texttt{vU}$x$\texttt{\_}\texttt{vN}$y$ cluster. The cluster dynamics model applied here differs from the model used in Ref.~\citenum{Cooper2023} because we do not include antisites.

Shown in Fig.~\ref{fig:middle_SA_stack} are the SA results for UN.
The uranium vacancy \texttt{vU01\_vN00} dominates the sensitivity 
for temperatures below $2000\text{K}$ as shown in Fig.~\ref{fig:middle_SA_stack}(a).
Interestingly, U diffusivity is not strongly dependent on the uranium interstitial \texttt{Ui01\_Ni00}, however,
we have found that in specific temperature, pressure, and irradiation regimes it is the dominant defect.
The sensitivity indices for  $D_\text{N}$ and $D_\text{Xe}$ are shown respectively in Figs.~\ref{fig:middle_SA_stack}(b) and \ref{fig:middle_SA_stack}(c).
For N diffusivity, the nitrogen interstitial \texttt{Ui00\_Ni01} defect is
almost 10 times more important than all the other defects
across the temperature range sampled. 
This is consistent with interstitials dominating nitrogen diffusion under nitrogen rich conditions.
Compare this to the results for Xe diffusion shown in Fig.~\ref{fig:middle_SA_stack}(c) which show that 
a number of defects contribute over 10\% to the model output.
All of those defects are linked to Xe diffusing by a vacancy mechanism, which is also the mechanism observed to dominate for most temperatures. 
UN corresponds to the perfect lattice, which impacts all defect energies and consequently appears in all sensitivities. It is possible that it represents a constant shift in all defect energies.
The sensitivity indices for U and N diffusivity do not vary strongly as the temperature is varied.
Across all the examined models, 
only a few defects make major contributions to the overall diffusivity in the diffusion
of the three species (U,N,Xe) we have examined.
 
\subsubsection{Optimization of {\tt CENTIPEDE} Parameters}
\begin{figure}[t]
\centering
\includegraphics[width = 8.5cm]{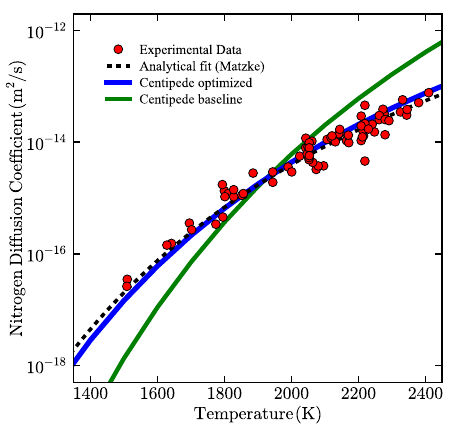}
\caption{\label{fig:genetic}
Nitrogen diffusion coefficient in UN as a function of temperature. The red points are experimental results (see Table~\ref{tab:table1}). The dashed black curve is the analytical fit by Matzke. The green curve is the baseline result from {\tt CENTIPEDE} and the blue curve is the {\tt CENTIPEDE} result after performing the genetic optimization procedure.
}
\end{figure}

The parameters in the UN cluster dynamics model implemented using {\tt CENTIPEDE} 
can be calibrated such that the quantities of interest agree with experimental results.
A detailed discussion of the specific physical parameters that are used in our model can be found in Ref.~\citenum{Cooper2023}.
However, the process of calibrating the
approximately $50$-dimensional parameter space
to experimental data by hand can be very time-consuming if not intractable. 
To accelerate the calibration process, 
we used a genetic optimization procedure to find the optimal set of parameters that produce {\tt CENTIPEDE} results that 
best match the experimental data for N diffusion. 
As a proof of principle, we perform the optimization on N diffusion only, 
and do not include U and Xe diffusion values in the optimization. 
A multi-objective optimization that includes diffusion for all of the species is a target of future work.

The genetic optimization procedure was implemented using a baseline set of  {\tt CENTIPEDE} parameter values obtained from a combination of electronic structure calculations, and molecular dynamics simulations. 
An error bound for each parameter was also assigned. 
Each set of parameter values is a solution to the optimization problem.
An initial set of solutions was generated by randomly sampling a value for each parameter using the assigned error bounds.
This initial set of solutions is called a generation.
Genetic optimization of the baseline parameters was performed by generating new sets of solutions (i.e., by generating new generations) using the best solutions from the previous generation. Each generation had 60 solutions and the optimization was performed for 5 generations. The procedure was performed across the temperature range $1300\text{K}-2500\text{K}$.

The results of the genetic optimization procedure are shown in Fig.~\ref{fig:genetic}. The diffusion values generated using the {\tt CENTIPEDE} parameters chosen as the baseline do not agree with the experimental results. However, after performing optimization in the high-dimensional parameter space, the {\tt CENTIPEDE} result is in strong agreement with the experimental data. The power of this result is that while the analytical result of Matzke (shown as dashed black curve) accurately captures the temperature dependence of N diffusion in the equilibrium regime, it cannot be used predict the diffusion behavior under different irradiation conditions and at different partial pressures. Compare this to the calibrated  {\tt CENTIPEDE} result which can be used to predict diffusion coefficients at  different system conditions. The calibrated cluster dynamics model is therefore more robust than simple analytical models for modeling diffusion under reactor conditions.
Overall, the defect parameters that change the most during the optimization procedure are kinetic parameters such as attempt frequencies and migration activation energies, while the enthalpic and entropic properties related to defect formation, in general, exhibit the least difference between the calibrated and baseline sets.
Moreover, the Centipede model predicts diffusion of not only N, which is relatively easy to measure experimentally, but also U and Xe, which are much harder to measure experimentally, especially under irradiation conditions. By validating and optimizing the {\tt CENTIPEDE}  model for N diffusion data, the trust in the {\tt CENTIPEDE}  predictions of U and Xe diffusion is also increased. Future work will investigate how these relations can be quantified using uncertainty measures.

\section{\label{sec:conc}Conclusions}

A data-driven workflow has been developed to predict diffusion coefficients in nuclear fuels.
The developed workflow has been shown to 
predict transport and thermodynamic properties of the nuclear fuels uranium oxide $\text{UO}_2$ and uranium nitride $\text{UN}$
with increased accuracy in comparison to previous models and methods.
We have specifically shown that using ML can reduce the predictive error in comparison to previously developed analytical models 
and to reduce the time it takes to develop diffusion models.
We have also shown how data-driven methods can be used to calibrate
complex mechanistic models for diffusion properties of nuclear fuels.
Machine learning models trained using small experimental datasets 
were expanded by developing and applying a data augmentation method.
This augmentation technique may be particularly useful in 
nuclear fuel development and qualification when large amounts of 
experimental results are not available or are difficult to obtain.
Sensitivity analysis methods have been applied
to determine the most important structural defects
within the mechanistic cluster dynamics models under different
reactor conditions.
This analysis can be used to 
improve model development and fuel analysis by giving information about what defect types should be targeted for further study using experiments and/or electronic structure calculations.

In future work, 
data-driven methods will be applied to
enhance the predictive capabilities of mechanistic models
for use in nuclear fuel qualification and reactor modeling.
Data-driven methods will also be used to generate complex multi-dimensional analytical functions with
enhanced transferability and interpretability in comparison to black-box ML models developed here.
Another important future focus will be to quantify the uncertainty in the developed ML models.

\section{Acknowledgments}
This work was supported by the U.S. Department of
Energy through the Los Alamos National Laboratory. Los
Alamos National Laboratory is operated by Triad National
Security, LLC, for the National Nuclear Security Administration of U.S. Department of Energy.
This research was supported by the Laboratory Directed Research and Development program of Los Alamos National Laboratory under project number 20220053DR. 
The computing resources used to perform this
research were provided by the LANL Institutional Computing Program.

\appendix \section{\label{app}Analytical Diffusion Models}

\noindent{\bf Hayes Model:}

The Hayes analytical model for U diffusion in UN in units of $\text{m}^2/\text{s}$ is:
\begin{equation}
D_\text{U}(T,P) = c_1 p^{c_2} e^{c_3/T}
\end{equation}
with $c_1 = 2.215\times10^{-15}$, $c_2 = 0.6414$, and $c_3 = -7989.3$,
where $p$ is the partial pressure of
nitrogen (atm) and $T$ is the temperature (K). 
\\

\noindent{\bf Matthews Model:} 

The Matthews analytical model for Xe diffusion in $\text{UO}_2$ in units of $\text{m}^2/\text{s}$ is:

\begin{equation}
D_\text{Xe}(T,\dot{F}) = \frac{c_1 e^{c_2/k_\text{B}T}}{c_3+c_4 e^{c_5/k_\text{B}T}} + c_6 e^{c_7/k_\text{B}T} \sqrt{\dot{F}}+ c_8\dot{F}
\end{equation}
with 
$c_1 = 2.216\times10^{-7}$, $c_2 = -3.26$,
$c_3 = 1.0$,
$c_4 = 29.03$,
$c_5 = -1.84\times10^{-4}$,
$c_6 = 2.821\times10^{-22}$,
$c_7 = -2.0$,
and
$c_8 = 8.5\times10^{-40}$
where $\dot{F} = 1.0\times10^{19}$ is the fission rate ($\text{fissions} / \text{m}^3 \, \text{s}$), $T$ is the temperature (K), and $k_\text{B}$ is the Boltzmann constant ($\text{eV}/\text{K}$). 
\\

\noindent{\bf Forsberg Model:}

The Forsberg analytical model for Xe diffusion in $\text{UO}_2$ in units of $\text{m}^2/\text{s}$ is:

\begin{equation}
D_\text{Xe}(T,\dot{F}) = \frac{v_g(T,\dot{F})D_\text{eff}(T,\dot{F})}{v_g(T,\dot{F})+ g(T,\dot{F})}
\end{equation}
with
\begin{align}
D_\text{eff}(T,\dot{F}) &= c_1 e^{c_2/T} + 4 c_3 e^{c_4/T} \sqrt{\dot{F}}+ 4 c_5 \dot{F},\\
v_g(T,\dot{F}) &= c_6 \pi l \dot{F} (c_7e^{c_8 T} + \delta)^2,\\
g(T,\dot{F}) &= 4 \pi c_7e^{c_8 T} (c_9/T -c_{10}) D_\text{eff}(T,\dot{F}),
\end{align}
where $\dot{F} = 1.72\times10^{19}$ is the fission rate ($\text{fissions} / \text{m}^3 \, \text{s}$)  and $T$ is the temperature (K).
The parameters are
$c_1 = 7.6\times10^{-10}$, 
$c_2 = -35247$,
$c_3 = 1.41\times10^{-25}$,
$c_4 = -13800$,
$c_5 = 2.0\times10^{-40}$,
$c_6 = 3.03$,
$l = 6.0\times10^{-6}$,
$c_7 = 1.453\times10^{-10}$,
$c_8 = 1.023\times10^{-3}$,
$\delta = 1.0\times10^{-9}$,
$c_9 = 1.52\times10^{27}$,
$c_{10} = 3.3\times10^{23}$.

\bibliographystyle{apsrev}

\begin{thebibliography}{64}
\bibitem[{not()}]{note1}
\bibinfo{note}{The Sturiale and DeCrescente (S\&D) diffusion values are
  sometimes noted as being too high due to the specific experimental technique
  that was used to obtain them. However, a change in temperature scales in the
  reported data from Celsius to Kelvin makes these values well aligned with the
  diffusion values measured by other sources. We therefore change the
  temperature scale in the S\&D data here.}
\end{thebibliography}

\end{document}